\documentclass[%
superscriptaddress,
preprint,
 amsmath,amssymb,
 aps,
]{revtex4-1}

\usepackage{graphicx}
\usepackage{dcolumn}
\usepackage{bm}
\usepackage{amsmath}
\usepackage{epsfig}
\usepackage{rotating}
\usepackage{enumerate}
\usepackage{xcolor}

\begin{document}

\title{Band gap engineering of MoS$_2$ upon compression}

\author{Miquel L\'opez-Su\'arez}
\affiliation{NiPS Laboratory, Dipartimento di Fisica e Geologia,
             Universit\`a degli Studi di Perugia, 
             06123 Perugia, Italy}
\email{miquel.lopez@nipslab.org}

\author{Igor Neri}
\affiliation{NiPS Laboratory, Dipartimento di Fisica e Geologia,
             Universit\`a degli Studi di Perugia, 
             06123 Perugia, Italy}
\affiliation{INFN Sezione di Perugia, via Pascoli,
             06123 Perugia, Italy}

\author{Riccardo Rurali}
\affiliation{Institut de Ci\`encia de Materials de Barcelona (ICMAB--CSIC)
             Campus de Bellaterra, 08193 Bellaterra, Barcelona, Spain}

\date{\today}

\begin{abstract}
Molybdenum disulfide (MoS$_2$) is a promising candidate  
for 2D nanoelectronic devices, that shows a direct band-gap for
monolayer structure. In this work we study the electronic structure of
MoS$_2$ upon both compressive and tensile strains with first-principles
density-functional calculations for different number of layers. 
The results show that the band-gap can be engineered for
experimentally attainable strains (i.e. $\pm 0.15$). However
compressive strain can result in bucking that can prevent the
use of large compressive strain.
We then studied the stability of the compression, calculating
the critical strain that results in the on-set of buckling
for free-standing nanoribbons of different lengths. The results
demonstrate that short structures, or few-layer MoS$_2$, show 
semi-conductor to metal transition upon compressive strain
without bucking.
\end{abstract}

\maketitle

\section{Introduction}

Molybdene disulfide, MoS$_2$, is a Transition Metal Dichalcogenide, TMD, with 
a hexagonal structure like graphene~\cite{WangNatureNano12}. As in 
the case of graphene, MoS$_2$\ can be exfoliated down to a single 
sheet~\cite{HaiACR14,LiuACSNano14} composed by one layer of Mo atoms 
stacked between two sulfide layers (see Fig.\ref{fig:struct}(a)).
Single layer MoS$_2$ is a direct gap semiconductor~\cite{MakPRL10} 
with $E_g = 1.9$~eV and high stiffness which makes it a very promising 
material for new nano electromechanical devices~\cite{BertolazziACSNano11}. 
Layered materials of this class are especially
amenable to bang-gap engineering upon strain, and monolayer and bilayer
MoS$_2$, in particular, have been shown to ultimately undergo a semiconductor-metal 
transition by means of mechanical strain~\cite{JohariACSNano12}. 
This transition occurs for tensile strains of around $\varepsilon=0.1$,
where $\varepsilon=\delta l/l$, with $l$ the original length of the structure.
The effect of compressive strain, however, has been thus far neglected.
The reason are twofold: (i)~experiments are naturally performed applying
a tensile strain; (ii)~compression is, at least in principle, a source
of structural instability in a two dimensional material, as it can 
result in buckling or other types of out-of-plane deformations~\cite{ZhangJMPS14}.

In this work we present first-principles electronic structure 
calculations of monolayer, bilayer, few-layer and bulk MoS$_2$
under axial and biaxial compressive strains of up to $\varepsilon=0.15$. We
explicitly address the stability of monolayer MoS$_2$ upon
compression, calculating the threshold strain beyond which
the accumulated elastic energy is relaxed through
buckling of the system. Additionally, we also extend the
known results of MoS$_2$ under tensile strain~\cite{JohariACSNano12,YunPRB12} 
considering mono-, few-layer and bulk MoS$_2$, showing that these systems
too can experience semiconductor-metal transition 
considering strains from $\varepsilon=0.08$ to $\varepsilon=0.12$, that can be achieved
in the experiments (i.e. $\varepsilon=0.23$)~\cite{CooperPRB13}.

\section{Methods}

\begin{figure}[t]
\begin{center}
\includegraphics[width=1.0\columnwidth]{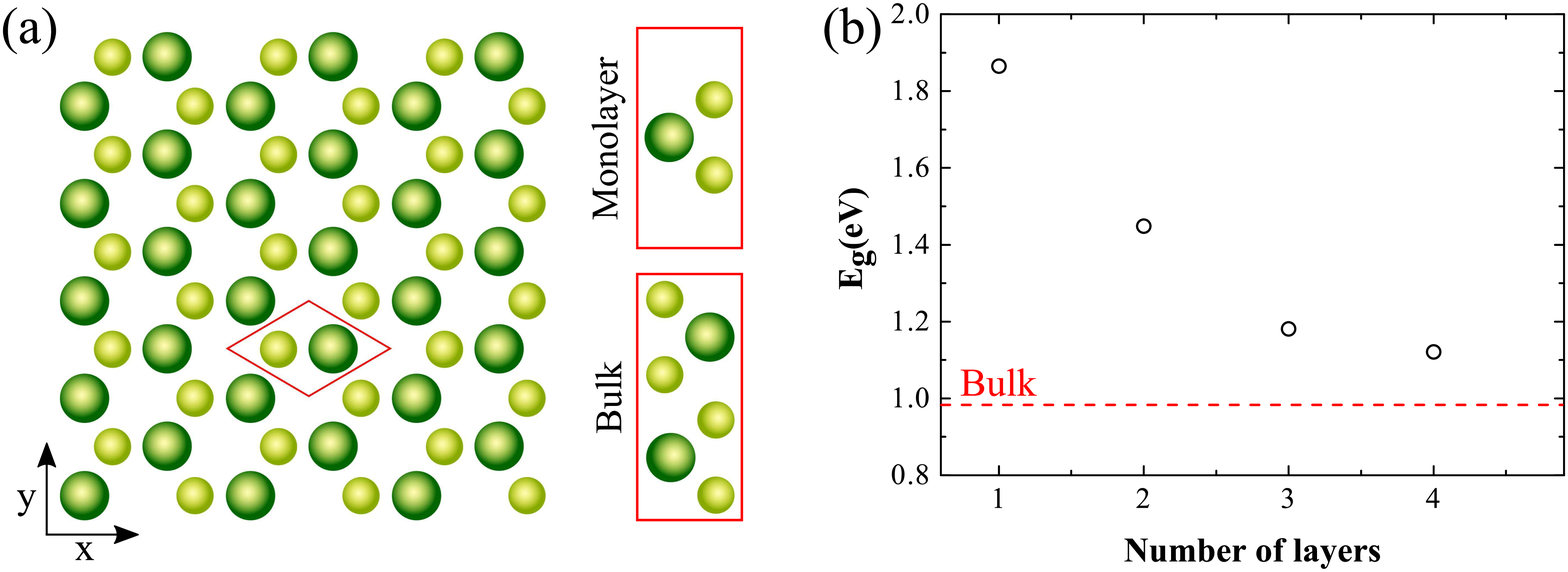}
\end{center}
\caption{(a) MoS$_2$ structure: red boxes highlight the primitive cells for top view (rhombus), and lateral views for mono layer and bulk (rectangles). (b) Energy gap for various number of layers (dots) and bulk (dashed line) MoS$_2$. }
\label{fig:struct}
\end{figure}

First principles calculations are carried out within Density
Functional Theory (DFT), as implemented in the {\sc Siesta} 
package~\cite{SolerJPCM02}. 
We use the Perdew-Burke-Eznerhof
parameterization of the Generalized Gradient Approximation 
(GGA)~\cite{PerdewPRL96} and an optimized double-$\zeta$ 
polarized basis set to expand the one-electron wave-function.
Core electrons are accounted for by means of norm-conserving
pseudopotentials. A converged grid of {\it k}-points 
to sample the Brillouin zone (the number of points and the direction
of the reciprocal space samples depend on the dimensionality and
cell size of the different systems studied, i.e. layered 2D
materials, bulk or nanotube) is used.
All structures are relaxed until all forces are lower than 
0.04~eV/\AA. The unit-cells of mono-layer and bulk MoS$_2$ are 
shown in Fig.~\ref{fig:struct}(a). The values for the lattice 
constant for the different structures considered in this work 
are listed in Table~\ref{tab:a}, which are in good agreement with 
the literature~\cite{AtacaJPCC11}. Strain is introduced to the system by 
deforming the unit-cell along the $x$ and $y$ axis for biaxial 
strain while deformation only along the $x$-axis is considered 
for uniaxial strain. In the case of uniaxial strain we study
both the case in which the unstrained lattice vector is and is not 
optimized. The former gives access to the Poisson's ratio, besides the
Young's modulus; the latter more closely models the situation
in which a MoS$_2$ layer is strongly adhered to a mismatched substrate
that compresses it only along one direction, but not along the other.
The band structures of all unstrained structures present an indirect 
band gap except that for mono-layer MoS$_2$ which exhibits a direct 
band gap placed at the K point, also in good agreement with previous
studies~\cite{MakPRL10}.
The energy band gap, $E_g$, as a function of the number of layers
is shown in Fig.~\ref{fig:struct}(b).

\section{Results}

\begin{figure}[t]
\begin{center}
\includegraphics[width=1.0\columnwidth]{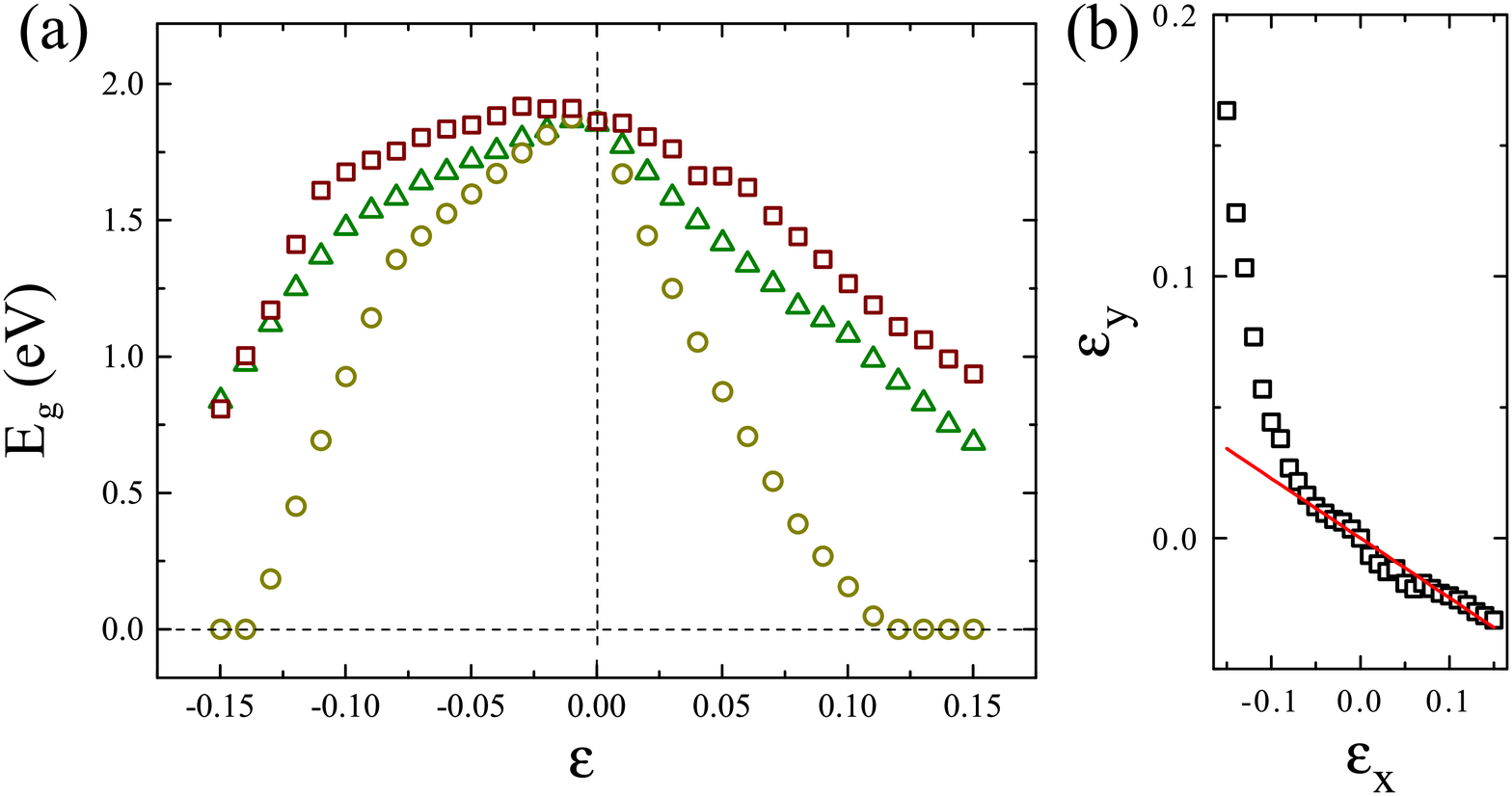}
\end{center}
\caption{(a) Dependence of the band gap of a monolayer of MoS$_2$ as 
         a function of the applied biaxial strain (circles) and
         uniaxial strain when the material is (squares) and is not
         (triangles) free to expand/contract in the perpendicular direction.
         (b) Induced strain along $y$ as a function of the applied
         strain along $x$ in the case of uniaxial strain of a free 
         standing MoS$_2$ monolayer (squares in panel (a)). 
         We found a negative ratio of transverse to axial strain resulting in a
         Poisson's ratio, $\nu$, of 0.23.
         }
\label{fig:Egap_vs_strain}
\end{figure}

The dependence of $E_g$ for monolayer MoS$_2$ with compressive and
tensile strain is shown in Fig.~\ref{fig:Egap_vs_strain}. Both tensile and
compressive strain produce a reduction of the band gap, 
regardless whether it is uniaxial or biaxial. In particular, 
biaxial compressive strain is the more effective way to tune the 
band gap and to ultimately drive a transition from semiconducting 
to metallic character, occurring at $\varepsilon=-0.14$, while the less 
effective method is compressive uniaxial strain, 
where $E_g$ shrinks of at most 1~eV for the largest values of 
the strain considered. Free standing monolayers --where 
upon uniaxial compression the sheet is free to transversely
expand (see Fig.~\ref{fig:Egap_vs_strain} (b))-- 
are slightly less sensitive to the applied strain, though
the differences become negligible at high compressions.

\begin{figure}[t]
\begin{center} 
\includegraphics[width=1.0\columnwidth]{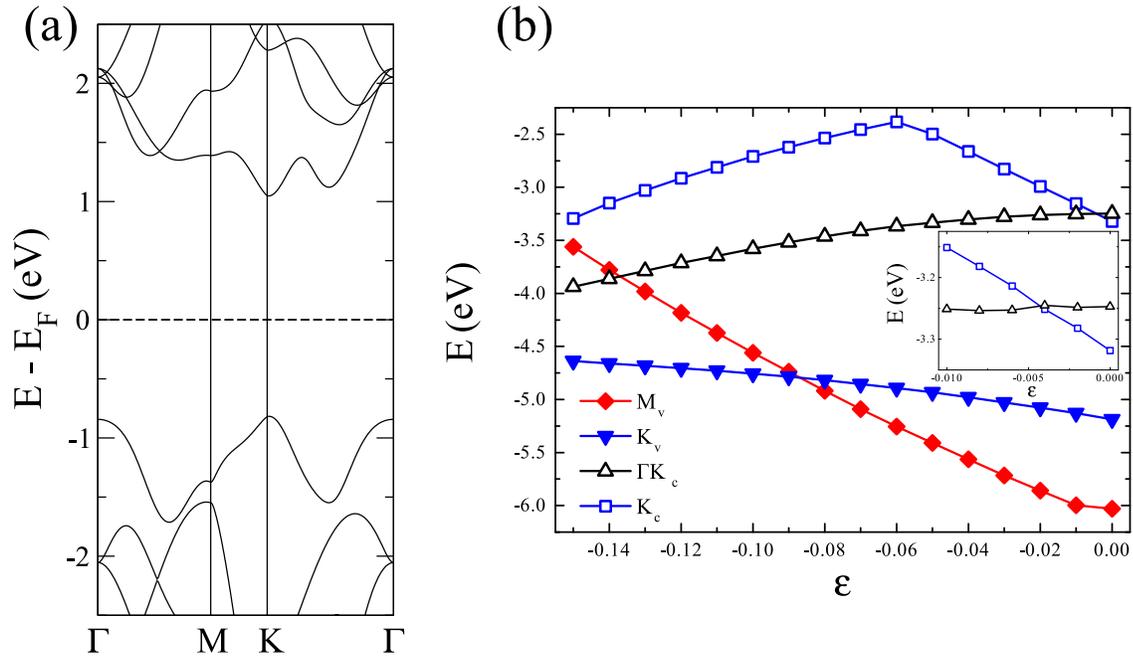}
\end{center}
\caption{(a)~Band-structure diagram of an unstrained MoS$_2$ monolayer.
         (b)~Dependence of the band-edge eigenvalues as a function
         of biaxial strain; inset: zoom of the low strain region
         where the band gap becomes indirect. 
         }
\label{fig:bands_eigs}
\end{figure}

A close inspection reveals a change of slope in the decrease
of the band gap as a function of biaxial strain around 
$\varepsilon=-0.08$. In order to understand this behavior we 
have tracked the dependence on the compressive strain of 
a few eigenvalues at high symmetry points (see the band-structure
diagram of Fig.~\ref{fig:bands_eigs}(a) for labeling).
It turns out that the valence band maximum at the $M$ point
increases much more quickly than the one at $K$ and for 
compressive strains larger than $\sim$~-0.08 it becomes the 
absolute maximum of the valence band. Therefore, the shrinking 
of band gap is determined by the pressure coefficient of the
valence band at $M$, while the larger pressure coefficient 
of the $K$ point takes over at larger compression.
A tiny compressive strain, on the other hand, is sufficient
to have the minimum of the conduction band at a point on the
$\Gamma-K$ path ($\Gamma K_{c}$), approximately equidistant from the two 
ends. This can be seen in the inset of Fig.~\ref{fig:bands_eigs}(b) 
where we have expanded the small strain region ($|\varepsilon|$
smaller than $0.01$).
Therefore, the band gap remains direct at the $K$ point only for
$-0.005 < \varepsilon < 0$; when $-0.08 < \varepsilon < -0.005$ the
gap is indirect because the minimum of the conduction is
along the $\Gamma-K$ path; finally, for $\varepsilon < -0.08$
is indirect between $M$ and $\Gamma-K$. The same behavior is present
also for the uniaxial strain, relaxed and not, even though the minimum
indirect gap is for different paths respect to the biaxial case.

These results indicate that the band gap of monolayer MoS$_2$ can
be engineered through compressive strain, similarly to what 
has been already shown with tensile strain. The response to strain
is of the same order in both cases. However, one
of the reasons that make compressive strain a less appealing 
way to engineer the band gap is that, at variance with tensile
strain, at a high enough compression the flat geometry becomes
unstable and buckling of the two-dimensional system is favored. 
While these buckled geometries can be useful for non-linear
energy harvesting of vibrational energy, as reported previously
by some of us~\cite{LopezSuarezPRB11,LopezSuarezMicroEng13,LopezSuarezNanotech14},
they are far to be ideal from the device design viewpoint of, say,
a field-effect transistor. Atom-thick graphene buckles even for 
very small strain values~\cite{LopezSuarezPRB11}, but previous reports hinted that 
for MoS$_2$ the flat geometry remains stable in a non-negligible
range of compression~\cite{MiquelPhD}. 

\begin{figure}[t]
\begin{center}
\includegraphics[width=1.0\columnwidth]{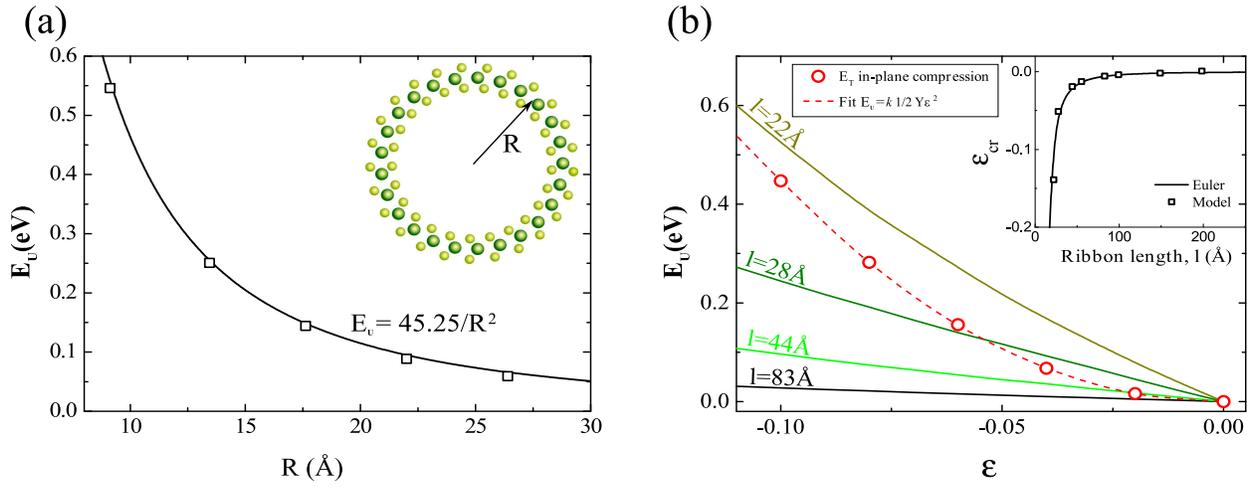}
\end{center}
\caption{(a) Curvature energy: energy per unit formula ($E_U$) of a MoS$_2$ nanotube
         as a function of the radius. (b) Comparisons between in-plane
         compression (red circles, dashed line is the fit) and
         buckling for different lengths, $l$, (continuous lines);
         inset: critical strains at which bucking is favored.
         }
\label{fig:mos2nts}
\end{figure}

In order to find the maximum compression that a ribbon of length
$l$ can support before buckling, we have compared the total energy 
under bending and under in-plane compression.
The energy per unit formula ($E_U$) of MoS$_2$ under bending has been calculated considering
nanotubes of different radius $R$.
The energy as function of the curvature radius varies as $1/R^2$, as
shown in Fig.~\ref{fig:mos2nts}(a). When a ribbon buckles the curvature
along its length is not constant and therefore the energy must be 
computed accordingly to the resulting curvature. In order to do so, 
we assume that the out-of-plane atomic displacements follow 
$u(x)=A \; \text{sin}(2 \pi/l x)$ and that the total length of the ribbon is 
constant and equal to its initial value, $l$.
The energy per unit formula of a MoS$_2$ as function of in-plane compression is shown in Fig.~\ref{fig:mos2nts}(b) 
(red circles), and can be approximated to $E_U=k 1/2 Y \varepsilon^2$, where $Y$
is the Young's modulus and $\varepsilon$ is the in-plane strain.

The critical strain, $\varepsilon_c$, is the strain at which buckling becomes more favorable 
than in-plane compression. This is shown in Fig.~\ref{fig:mos2nts}(b) 
as crossing points between the red line and the different
lines corresponding to the bending energy for different ribbon lengths.
The model agrees well with the prediction of Euler elasticity theory, 
black continuous line in Fig.~\ref{fig:mos2nts}(b), to be compared
with the black squares, i.e. the critical strains obtained from the data.

\begin{figure}[t]
\begin{center}
\includegraphics[width=1.0\columnwidth]{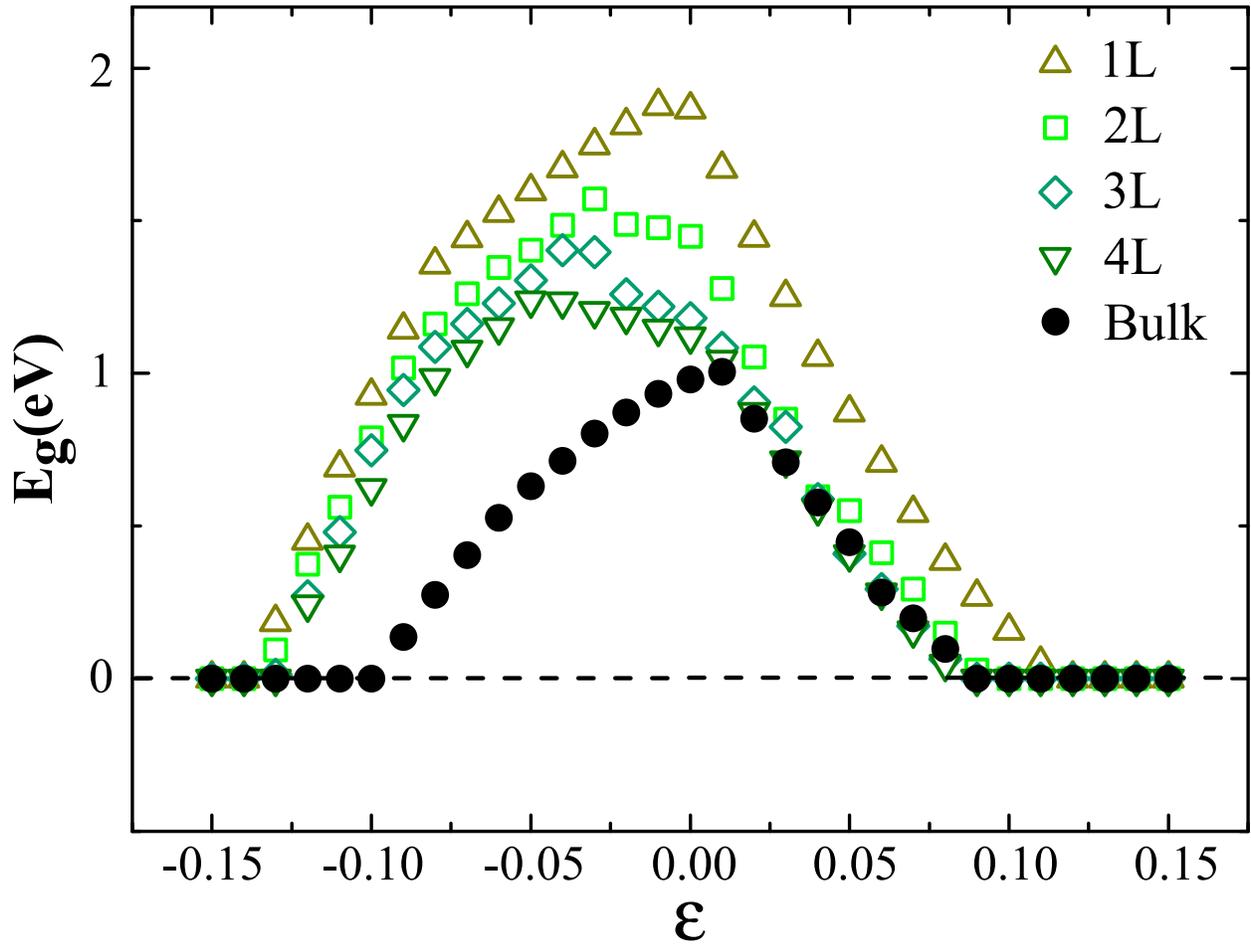}
\end{center}
\caption{Energy gap for mono, few-layer, and bulk MoS$_2$ as function of biaxial strain. For bulk material metal transition occurs at lower compressive strain compared to mono and few-layer. For stretching strains the transition appears at $\varepsilon\sim$~0.08 for bulk and few-layer MoS$_2$.
         }
\label{fig:mlayer}
\end{figure}

It seems clear that, for monolayer MoS$_2$, compressions larger than 
$-0.05$ can be achieved without buckling only for nanoribbons shorter 
than $3$~nm. Ribbons with more attainable dimensions, i.e. $l>100$~nm, 
buckle for $\varepsilon < -0.001$ which effectively prevent the modulation of 
the band-gap by compressive strain.
Nonetheless, the bending energy increases for thicker 
structures, i.e. bi-layer, tri-layer, approaching infinite for 
bulk materials. The critical strain $\varepsilon_c$ also increases, 
widening the range of compressive strains can be attained without 
inducing buckling, even in systems of longer lengths. For this 
reason, we have also calculated the response to strain of few-layer 
and bulk MoS$_2$, finding a qualitative similar behavior (see 
Fig.~\ref{fig:mlayer}). Our results show not only a decrease of 
the energy band gap for unstrained few-layer MoS$_2$ 
(reaching $E_g=$~1~eV for bulk ), but also a 
the possibility of achieving a semiconductor-to-metal transition 
for high applied compressive strains in all cases. Noteworthy,
the transition occurs even for slightly lower strain values, 
both compressive and tensile, for increasing number of layers.
As a final remark one should note that, while predictions of the
pressure coefficients based on DFT calculations are very reliable,
the band-gaps are notoriously underestimated. This means that 
the slopes of the curves in Fig.~\ref{fig:Egap_vs_strain} and 
\ref{fig:mlayer} are accurate, but closing the band-gap likely 
requires larger strains. 

\section{Conclusions}

In conclusion, we have shown that both tensile and compressive strain
result in band gap engineering of  monolayer, few-layer and bulk MoS$_2$.
A transition from semiconductor to metal can be achieved for compressions 
of the order of $\varepsilon = 0.13$ for monolayer MoS$_2$ under biaxial 
strain, while for bulk MoS$_2$ this value is reduced to $0.10$.
We have also computed the maximum compression that a MoS$_2$ monolayer can stand
without favoring the onset of buckling instabilities, thus assessing
within which range compressive strain can be used to tailor the electronic
properties of a flat MoS$_2$ sheet. From the latter calculations we show that
the maximum dimensions for single layer MoS$_2$ able to show a metal transition 
before bucking are impracticable (i.e. maximum $3$~nm). However the bucking 
can be avoided considering few-layer suspended structures, that also show semiconductor
to metal transition, or the confinement of the MoS$_2$ to a substrate, which 
can also delay the emergence of buckling.

\begin{acknowledgments}
MLS and IN gratefully acknowledge financial support from the European 
Commission (FPVII, Grant agreement no: 318287, LANDAUER and Grant 
agreement no: 611004, ICT- Energy). 
RR acknowledges the Severo Ochoa Centres of Excellence Program under Grant
SEV-2015-0496, funding under contracts Nos. FEDER-MAT2013-40581-P
of the Ministerio de Econom\'ia y Competitividad (MINECO)
and grant 2014 SGR 301 of the Generalitat de Catalunya.
\end{acknowledgments}



\clearpage

\bibliography{bibliography.bib}

\begin{thebibliography}{16}%
\makeatletter
\providecommand \@ifxundefined [1]{%
 \@ifx{#1\undefined}
}%
\providecommand \@ifnum [1]{%
 \ifnum #1\expandafter \@firstoftwo
 \else \expandafter \@secondoftwo
 \fi
}%
\providecommand \@ifx [1]{%
 \ifx #1\expandafter \@firstoftwo
 \else \expandafter \@secondoftwo
 \fi
}%
\providecommand \natexlab [1]{#1}%
\providecommand \enquote  [1]{``#1''}%
\providecommand \bibnamefont  [1]{#1}%
\providecommand \bibfnamefont [1]{#1}%
\providecommand \citenamefont [1]{#1}%
\providecommand \href@noop [0]{\@secondoftwo}%
\providecommand \href [0]{\begingroup \@sanitize@url \@href}%
\providecommand \@href[1]{\@@startlink{#1}\@@href}%
\providecommand \@@href[1]{\endgroup#1\@@endlink}%
\providecommand \@sanitize@url [0]{\catcode `\\12\catcode `\$12\catcode
  `\&12\catcode `\#12\catcode `\^12\catcode `\_12\catcode `\%12\relax}%
\providecommand \@@startlink[1]{}%
\providecommand \@@endlink[0]{}%
\providecommand \url  [0]{\begingroup\@sanitize@url \@url }%
\providecommand \@url [1]{\endgroup\@href {#1}{\urlprefix }}%
\providecommand \urlprefix  [0]{URL }%
\providecommand \Eprint [0]{\href }%
\providecommand \doibase [0]{http://dx.doi.org/}%
\providecommand \selectlanguage [0]{\@gobble}%
\providecommand \bibinfo  [0]{\@secondoftwo}%
\providecommand \bibfield  [0]{\@secondoftwo}%
\providecommand \translation [1]{[#1]}%
\providecommand \BibitemOpen [0]{}%
\providecommand \bibitemStop [0]{}%
\providecommand \bibitemNoStop [0]{.\EOS\space}%
\providecommand \EOS [0]{\spacefactor3000\relax}%
\providecommand \BibitemShut  [1]{\csname bibitem#1\endcsname}%
\let\auto@bib@innerbib\@empty
\bibitem [{\citenamefont {Wang}\ \emph {et~al.}(2012)\citenamefont {Wang},
  \citenamefont {Kalantar-Zadeh}, \citenamefont {Kis}, \citenamefont
  {Coleman},\ and\ \citenamefont {Strano}}]{WangNatureNano12}%
  \BibitemOpen
  \bibfield  {author} {\bibinfo {author} {\bibfnamefont {Q.~H.}\ \bibnamefont
  {Wang}}, \bibinfo {author} {\bibfnamefont {K.}~\bibnamefont
  {Kalantar-Zadeh}}, \bibinfo {author} {\bibfnamefont {A.}~\bibnamefont {Kis}},
  \bibinfo {author} {\bibfnamefont {J.~N.}\ \bibnamefont {Coleman}}, \ and\
  \bibinfo {author} {\bibfnamefont {M.~S.}\ \bibnamefont {Strano}},\
  }\href@noop {} {\bibfield  {journal} {\bibinfo  {journal} {Nat. Nanotech.}\
  }\textbf {\bibinfo {volume} {7}},\ \bibinfo {pages} {699} (\bibinfo {year}
  {2012})}\BibitemShut {NoStop}%
\bibitem [{\citenamefont {Li}\ \emph {et~al.}(2014)\citenamefont {Li},
  \citenamefont {Wu}, \citenamefont {Yin},\ and\ \citenamefont
  {Zhang}}]{HaiACR14}%
  \BibitemOpen
  \bibfield  {author} {\bibinfo {author} {\bibfnamefont {H.}~\bibnamefont
  {Li}}, \bibinfo {author} {\bibfnamefont {J.}~\bibnamefont {Wu}}, \bibinfo
  {author} {\bibfnamefont {Z.}~\bibnamefont {Yin}}, \ and\ \bibinfo {author}
  {\bibfnamefont {H.}~\bibnamefont {Zhang}},\ }\href@noop {} {\bibfield
  {journal} {\bibinfo  {journal} {Acc. Chem. Res.}\ }\textbf {\bibinfo {volume}
  {47}},\ \bibinfo {pages} {1067} (\bibinfo {year} {2014})}\BibitemShut
  {NoStop}%
\bibitem [{\citenamefont {Liu}\ \emph {et~al.}(2014)\citenamefont {Liu},
  \citenamefont {Kim}, \citenamefont {Kim}, \citenamefont {Ye}, \citenamefont
  {Kim},\ and\ \citenamefont {Lee}}]{LiuACSNano14}%
  \BibitemOpen
  \bibfield  {author} {\bibinfo {author} {\bibfnamefont {N.}~\bibnamefont
  {Liu}}, \bibinfo {author} {\bibfnamefont {P.}~\bibnamefont {Kim}}, \bibinfo
  {author} {\bibfnamefont {J.~H.}\ \bibnamefont {Kim}}, \bibinfo {author}
  {\bibfnamefont {J.~H.}\ \bibnamefont {Ye}}, \bibinfo {author} {\bibfnamefont
  {S.}~\bibnamefont {Kim}}, \ and\ \bibinfo {author} {\bibfnamefont {C.~J.}\
  \bibnamefont {Lee}},\ }\href@noop {} {\bibfield  {journal} {\bibinfo
  {journal} {ACS Nano}\ }\textbf {\bibinfo {volume} {8}},\ \bibinfo {pages}
  {6902} (\bibinfo {year} {2014})}\BibitemShut {NoStop}%
\bibitem [{\citenamefont {Mak}\ \emph {et~al.}(2010)\citenamefont {Mak},
  \citenamefont {Lee}, \citenamefont {Hone}, \citenamefont {Shan},\ and\
  \citenamefont {Heinz}}]{MakPRL10}%
  \BibitemOpen
  \bibfield  {author} {\bibinfo {author} {\bibfnamefont {K.~F.}\ \bibnamefont
  {Mak}}, \bibinfo {author} {\bibfnamefont {C.}~\bibnamefont {Lee}}, \bibinfo
  {author} {\bibfnamefont {J.}~\bibnamefont {Hone}}, \bibinfo {author}
  {\bibfnamefont {J.}~\bibnamefont {Shan}}, \ and\ \bibinfo {author}
  {\bibfnamefont {T.~F.}\ \bibnamefont {Heinz}},\ }\href@noop {} {\bibfield
  {journal} {\bibinfo  {journal} {Phys. Rev. Lett.}\ }\textbf {\bibinfo
  {volume} {105}},\ \bibinfo {pages} {136805} (\bibinfo {year}
  {2010})}\BibitemShut {NoStop}%
\bibitem [{\citenamefont {Bertolazzi}\ \emph {et~al.}(2011)\citenamefont
  {Bertolazzi}, \citenamefont {Brivio},\ and\ \citenamefont
  {Kis}}]{BertolazziACSNano11}%
  \BibitemOpen
  \bibfield  {author} {\bibinfo {author} {\bibfnamefont {S.}~\bibnamefont
  {Bertolazzi}}, \bibinfo {author} {\bibfnamefont {J.}~\bibnamefont {Brivio}},
  \ and\ \bibinfo {author} {\bibfnamefont {A.}~\bibnamefont {Kis}},\
  }\href@noop {} {\bibfield  {journal} {\bibinfo  {journal} {ACS Nano}\
  }\textbf {\bibinfo {volume} {5}},\ \bibinfo {pages} {9703} (\bibinfo {year}
  {2011})}\BibitemShut {NoStop}%
\bibitem [{\citenamefont {Johari}\ and\ \citenamefont
  {Shenoy}(2012)}]{JohariACSNano12}%
  \BibitemOpen
  \bibfield  {author} {\bibinfo {author} {\bibfnamefont {P.}~\bibnamefont
  {Johari}}\ and\ \bibinfo {author} {\bibfnamefont {V.~B.}\ \bibnamefont
  {Shenoy}},\ }\href@noop {} {\bibfield  {journal} {\bibinfo  {journal} {ACS
  Nano}\ }\textbf {\bibinfo {volume} {6}},\ \bibinfo {pages} {5449} (\bibinfo
  {year} {2012})}\BibitemShut {NoStop}%
\bibitem [{\citenamefont {Zhang}\ and\ \citenamefont
  {Arroyo}(2014)}]{ZhangJMPS14}%
  \BibitemOpen
  \bibfield  {author} {\bibinfo {author} {\bibfnamefont {K.}~\bibnamefont
  {Zhang}}\ and\ \bibinfo {author} {\bibfnamefont {M.}~\bibnamefont {Arroyo}},\
  }\href@noop {} {\bibfield  {journal} {\bibinfo  {journal} {J. Mech. Phys.
  Solids}\ }\textbf {\bibinfo {volume} {72}},\ \bibinfo {pages} {61} (\bibinfo
  {year} {2014})}\BibitemShut {NoStop}%
\bibitem [{\citenamefont {Yun}\ \emph {et~al.}(2012)\citenamefont {Yun},
  \citenamefont {Han}, \citenamefont {Hong}, \citenamefont {Kim},\ and\
  \citenamefont {Lee}}]{YunPRB12}%
  \BibitemOpen
  \bibfield  {author} {\bibinfo {author} {\bibfnamefont {W.~S.}\ \bibnamefont
  {Yun}}, \bibinfo {author} {\bibfnamefont {S.~W.}\ \bibnamefont {Han}},
  \bibinfo {author} {\bibfnamefont {S.~C.}\ \bibnamefont {Hong}}, \bibinfo
  {author} {\bibfnamefont {I.~G.}\ \bibnamefont {Kim}}, \ and\ \bibinfo
  {author} {\bibfnamefont {J.~D.}\ \bibnamefont {Lee}},\ }\href@noop {}
  {\bibfield  {journal} {\bibinfo  {journal} {Phys. Rev. B}\ }\textbf {\bibinfo
  {volume} {85}},\ \bibinfo {pages} {033305} (\bibinfo {year}
  {2012})}\BibitemShut {NoStop}%
\bibitem [{\citenamefont {Cooper}\ \emph {et~al.}(2013)\citenamefont {Cooper},
  \citenamefont {Lee}, \citenamefont {Marianetti}, \citenamefont {Wei},
  \citenamefont {Hone},\ and\ \citenamefont {Kysar}}]{CooperPRB13}%
  \BibitemOpen
  \bibfield  {author} {\bibinfo {author} {\bibfnamefont {R.~C.}\ \bibnamefont
  {Cooper}}, \bibinfo {author} {\bibfnamefont {C.}~\bibnamefont {Lee}},
  \bibinfo {author} {\bibfnamefont {C.~A.}\ \bibnamefont {Marianetti}},
  \bibinfo {author} {\bibfnamefont {X.}~\bibnamefont {Wei}}, \bibinfo {author}
  {\bibfnamefont {J.}~\bibnamefont {Hone}}, \ and\ \bibinfo {author}
  {\bibfnamefont {J.~W.}\ \bibnamefont {Kysar}},\ }\href@noop {} {\bibfield
  {journal} {\bibinfo  {journal} {Phys. Rev. B}\ }\textbf {\bibinfo {volume}
  {87}},\ \bibinfo {pages} {035423} (\bibinfo {year} {2013})}\BibitemShut
  {NoStop}%
\bibitem [{\citenamefont {Soler}\ \emph {et~al.}(2002)\citenamefont {Soler},
  \citenamefont {Artacho}, \citenamefont {Gale}, \citenamefont {Garc\'{i}a},
  \citenamefont {Junquera}, \citenamefont {Ordej\'{o}n},\ and\ \citenamefont
  {S\'{a}nchez-Portal}}]{SolerJPCM02}%
  \BibitemOpen
  \bibfield  {author} {\bibinfo {author} {\bibfnamefont {J.~M.}\ \bibnamefont
  {Soler}}, \bibinfo {author} {\bibfnamefont {E.}~\bibnamefont {Artacho}},
  \bibinfo {author} {\bibfnamefont {J.~D.}\ \bibnamefont {Gale}}, \bibinfo
  {author} {\bibfnamefont {A.}~\bibnamefont {Garc\'{i}a}}, \bibinfo {author}
  {\bibfnamefont {J.}~\bibnamefont {Junquera}}, \bibinfo {author}
  {\bibfnamefont {P.}~\bibnamefont {Ordej\'{o}n}}, \ and\ \bibinfo {author}
  {\bibfnamefont {D.}~\bibnamefont {S\'{a}nchez-Portal}},\ }\href@noop {}
  {\bibfield  {journal} {\bibinfo  {journal} {J. Phys.: Condens. Matter}\
  }\textbf {\bibinfo {volume} {14}},\ \bibinfo {pages} {2745} (\bibinfo {year}
  {2002})}\BibitemShut {NoStop}%
\bibitem [{\citenamefont {Perdew}\ \emph {et~al.}(1996)\citenamefont {Perdew},
  \citenamefont {Burke},\ and\ \citenamefont {Ernzerhof}}]{PerdewPRL96}%
  \BibitemOpen
  \bibfield  {author} {\bibinfo {author} {\bibfnamefont {J.~P.}\ \bibnamefont
  {Perdew}}, \bibinfo {author} {\bibfnamefont {K.}~\bibnamefont {Burke}}, \
  and\ \bibinfo {author} {\bibfnamefont {M.}~\bibnamefont {Ernzerhof}},\
  }\href@noop {} {\bibfield  {journal} {\bibinfo  {journal} {Phys. Rev. Lett.}\
  }\textbf {\bibinfo {volume} {77}},\ \bibinfo {pages} {3865} (\bibinfo {year}
  {1996})}\BibitemShut {NoStop}%
\bibitem [{\citenamefont {Ataca}\ \emph {et~al.}(2011)\citenamefont {Ataca},
  \citenamefont {Topsakal}, \citenamefont {Akturk},\ and\ \citenamefont
  {Ciraci}}]{AtacaJPCC11}%
  \BibitemOpen
  \bibfield  {author} {\bibinfo {author} {\bibfnamefont {C.}~\bibnamefont
  {Ataca}}, \bibinfo {author} {\bibfnamefont {M.}~\bibnamefont {Topsakal}},
  \bibinfo {author} {\bibfnamefont {E.}~\bibnamefont {Akturk}}, \ and\ \bibinfo
  {author} {\bibfnamefont {S.}~\bibnamefont {Ciraci}},\ }\href@noop {}
  {\bibfield  {journal} {\bibinfo  {journal} {J. Phys. Chem. C}\ }\textbf
  {\bibinfo {volume} {115}},\ \bibinfo {pages} {16354} (\bibinfo {year}
  {2011})}\BibitemShut {NoStop}%
\bibitem [{\citenamefont {L\'opez-Su\'arez}\ \emph {et~al.}(2011)\citenamefont
  {L\'opez-Su\'arez}, \citenamefont {Rurali}, \citenamefont {Gammaitoni},\ and\
  \citenamefont {Abadal}}]{LopezSuarezPRB11}%
  \BibitemOpen
  \bibfield  {author} {\bibinfo {author} {\bibfnamefont {M.}~\bibnamefont
  {L\'opez-Su\'arez}}, \bibinfo {author} {\bibfnamefont {R.}~\bibnamefont
  {Rurali}}, \bibinfo {author} {\bibfnamefont {L.}~\bibnamefont {Gammaitoni}},
  \ and\ \bibinfo {author} {\bibfnamefont {G.}~\bibnamefont {Abadal}},\
  }\href@noop {} {\bibfield  {journal} {\bibinfo  {journal} {Phys. Rev. B}\
  }\textbf {\bibinfo {volume} {84}},\ \bibinfo {pages} {161401} (\bibinfo
  {year} {2011})}\BibitemShut {NoStop}%
\bibitem [{\citenamefont {L{\'o}pez-Su{\'a}rez}\ \emph
  {et~al.}(2013)\citenamefont {L{\'o}pez-Su{\'a}rez}, \citenamefont {Rurali},\
  and\ \citenamefont {Abadal}}]{LopezSuarezMicroEng13}%
  \BibitemOpen
  \bibfield  {author} {\bibinfo {author} {\bibfnamefont {M.}~\bibnamefont
  {L{\'o}pez-Su{\'a}rez}}, \bibinfo {author} {\bibfnamefont {R.}~\bibnamefont
  {Rurali}}, \ and\ \bibinfo {author} {\bibfnamefont {G.}~\bibnamefont
  {Abadal}},\ }\href@noop {} {\bibfield  {journal} {\bibinfo  {journal}
  {Microelectronic Engineering}\ }\textbf {\bibinfo {volume} {111}},\ \bibinfo
  {pages} {122} (\bibinfo {year} {2013})}\BibitemShut {NoStop}%
\bibitem [{\citenamefont {L\'opez-Su\'arez}\ \emph {et~al.}(2014)\citenamefont
  {L\'opez-Su\'arez}, \citenamefont {Pruneda}, \citenamefont {Abadal},\ and\
  \citenamefont {Rurali}}]{LopezSuarezNanotech14}%
  \BibitemOpen
  \bibfield  {author} {\bibinfo {author} {\bibfnamefont {M.}~\bibnamefont
  {L\'opez-Su\'arez}}, \bibinfo {author} {\bibfnamefont {M.}~\bibnamefont
  {Pruneda}}, \bibinfo {author} {\bibfnamefont {G.}~\bibnamefont {Abadal}}, \
  and\ \bibinfo {author} {\bibfnamefont {R.}~\bibnamefont {Rurali}},\
  }\href@noop {} {\bibfield  {journal} {\bibinfo  {journal} {Nanotechnology}\
  }\textbf {\bibinfo {volume} {25}},\ \bibinfo {pages} {175401} (\bibinfo
  {year} {2014})}\BibitemShut {NoStop}%
\bibitem [{Miq()}]{MiquelPhD}%
  \BibitemOpen
  \href@noop {} {}\bibinfo {howpublished} {M. L\'opez-Su\'arez, {\em Energy
  harvesting from the microscale to the nanoscale}. PhD Thesis, 2015.
  http://hdl.handle.net/10803/283731}\BibitemShut {NoStop}%
\end{thebibliography}%

\clearpage

\begin{table}
\caption{Values for the lattice constant for different number of layers}\label{tab:a}
\begin{tabular}{c|c|c|c|c|c}
 & 1L & 2L & 3L & 4L & Bulk \\ 
\hline 
a~(\AA) & 3.18 & 3.20 & 3.20 & 3.20 & 3.20 \\ 
\end{tabular} 
\end{table}

\clearpage

\clearpage

\clearpage

\clearpage

\end{document}